# Confinement-Induced Nonlocality and Optical Nonlinearity of Transdimensional Titanium Nitride in the Epsilon-Near-Zero Region

Fan-Ting Tseng,[1] I-Hung Ho,[1] Ting-Jui Kuo,[1] Shangjr Gwo,[2] Igor V. Bondarev,[3] and Hyeyoung Ahn[1*]

[1]Department of Photonics, College of Electrical and Computer Engineering, National Yang Ming Chiao Tung University, Hsinchu 30010, Taiwan

[2]Department of Physics, National Tsing-Hua University, Hsinchu 30013, Taiwan

[3]Department of Mathematics and Physics, North Carolina Central University, Durham, NC 27707, USA

**ABSTRACT.** Ultrathin plasmonic films that approach the transdimensional (TD) thickness limit provide a promising route for light–matter interaction control and manipulation, yet their nonlinear optical response near the epsilon-near-zero (ENZ) condition remains poorly understood. Here, we report the strongly enhanced optical nonlinearity for their typical representative—high-quality TiN epitaxial films with thicknesses down to a few nanometers. Systematic Z-scan measurements reveal a pronounced increase in nonlinear absorption with decreasing thickness. Especially in the ENZ spectral region, the TD TiN films exhibit nearly two orders of magnitude stronger nonlinear absorption over a broad range of incidence angles as compared to conventional thin films. The enhanced nonlinear absorption observed is well described by a nonlinear nonlocal electromagnetic response model that accounts for electron confinement effects unique to the TD plasmonic systems. Comparison with $Ti_{1-x}Al_xN$ highlights the necessity of low-loss ENZ response for nonlinear enhancement. These findings identify TiN and similar TD plasmonic systems as a robust refractory platform for exploiting ENZ-mediated nonlinear processes in ultrathin photonic material structures.

**Introduction**

Plasmonic metals play a central role in modern nanophotonics due to their ability to strongly confine and manipulate light at subwavelength scales through excitation of surface plasmons[1-3]. Moreover, the intense electromagnetic (EM) field enhancement near metallic interfaces enables efficient light-matter interactions far beyond the diffraction limit, thus facilitating nonlinear optical processes such as high harmonic generation, four-wave mixing, and optical Kerr effect[4-6]. However, the intrinsic nonlinearities of conventional metals are often weak and obscured by high optical losses[7,8]. Therefore, it is crucial to develop effective strategies to enhance the nonlinear optical response of plasmonic metals, particularly under low excitation intensities, to enable their practical implementation in integrated nanophotonic devices.

Recently, a promising approach to overcoming these limitations has emerged through the exploration of metals operating within their epsilon-near-zero (ENZ) region[9-15]. In this spectral domain, the real part of the permittivity ($\varepsilon_1$) of the ENZ materials approaches zero as the angular frequency approaches the plasma frequency ($\omega_p$). The resulting large refractive index modulation within the ENZ region leads to strong enhancement of the local electric field, thus amplifying nonlinear optical effects. ENZ phenomena have been explored in doped semiconductors and metamaterials[11-15], and several orders of magnitude enhancement in optical nonlinearity have been reported for transparent conducting oxides (TCOs), such as indium tin oxide (ITO), near their ENZ wavelengths[16-20]. Meanwhile, the excitation of ENZ modes is strongly correlated with film thickness as such modes can exist within a specific thickness range ($d$) approximately satisfying $d \lesssim \lambda_p/50$, where $\lambda_p = 2\pi c/\omega_p$[12-16]. Therefore, most ENZ effects in TCOs are limited to the near-infrared (NIR) spectral range and rely on extrinsic doping control. In contrast, metals possess plasma wavelengths in the visible range ($\lambda_p \approx 520$ nm for gold), and ENZ behavior in these materials typically emerges only when the films are ultrathin with nanoscale thicknesses[21,22]. However, as the film thickness decreases, the high optical loss in metals further increases as well, resulting in strong plasmon attenuation and shortened propagation lengths[23]. Therefore, the manifestation of ENZ-driven nonlinearities in metals crucially depends on achieving high crystalline quality with precisely controlled thickness down to a few nanometers.

Recently, ultrathin finite-thickness metallic films have attracted a great deal of interest with regard to transdimensional (TD) plasmonic materials[24-42]. The term "transdimensional" refers to the transitional range of thicknesses—a regime that is neither three (3D) nor two (2D) dimensional but rather something in between, turning into 2D for atomically thin monolayers and thus offering an opportunity to study a previously unexplored quasi-continuous 3D-to-2D transition (by monolayer number reduction) to improve material functionalities[24,25]. In this regime, the strong vertical quantum confinement makes the linear electromagnetic (EM) response of the film nonlocal (spatially dispersive), and the degree of nonlocality can be controlled by the film thickness[13,24]. As a result, one can probe the fundamental properties of light-matter interactions as they evolve from the 3D case of bulk materials to the 2D case of atomically thin (quantum confined) films of the same chemical composition[25]. This makes plasmonic TD materials indispensable for studies of the nonlocal light-matter interactions at the nanoscale[26-42], where they exhibit an extraordinary tailorability of their EM properties, including the capabilities of active control and tuning of their EM response by adjusting their thickness and thus enabling new and unique light-matter coupling phenomena[30,31].

The linear EM response nonlocality of TD plasmonic materials was first predicted theoretically[24] and then confirmed experimentally as a remarkable intrinsic property of planar TD metallic nanostructures[26-29]. It has been shown to enable a variety of new quantum phenomena in TD plasmonic film systems.[30,31] They are the thickness-controlled plasma frequency red shift[26,32], the low-temperature plasma frequency dropoff[28], the surface plasmon mode degeneracy lifting[12,13], a series of quantum-optical[33], nonlocal magneto-optical[34] and anisotropic quantum effects[35-38], including electronic transitions that are normally forbidden[39] and quantum metal-insulator transitions[27,40], as well as extraordinary thermal emission[29,41] and optical reflection properties[42].

In this work, we explore the nonlinear optical properties in the ENZ domain for thin and ultrathin (under 10 nm in thickness) films of titanium nitride (TiN), a typical representative of the transition metal nitride family[7]. Ultrathin TiN films of controlled thickness represent a new class of TD plasmonic materials[30,31], which is known for excellent optical and electrical properties close to those of noble metals but with a significant advantage of having exceptional chemical and thermal stability (melting point ~3000 °C) that are necessary to ensure a robust nonlinear response and compatibility with semiconductor fabrication process. Recent experimental measurements of TiN optical nonlinearities have revealed signatures of the ENZ effect[43-46], though primarily in TiN nanoparticle composites of quite large overall size (~500 nm)[45,46]. Here, on the contrary, we focus on the optical nonlinearities associated with the confinement-induced nonlocal EM response of TiN films with thicknesses in the range 2 to 50 nm that includes the TD thickness regime. While the growth of ultrathin films of noble metals is limited by the percolation threshold (~6 nm for gold[47,48]), ultrathin TiN films of thickness down to 2 nm with exceptional single-crystalline quality and precise thickness control can be successfully fabricated via nitrogen plasma-assisted molecular beam epitaxy (PA-MBE)[49,50]. Here, we use such TD films to study the interrelation between the confinement-induced nonlocality and thickness-dependent nonlinear optical absorption processes. Additionally, we explore the influence of the optical losses on nonlinear optical absorption using Al-doped $Ti_{1-x}Al_xN$ films, where the losses are systematically tuned by varying the aluminum composition[51]. To understand and explain remarkably unusual angular and thickness dependences of the nonlinear AO coefficient ($\beta$), we generalize the nonlocal linear EM response theory of TD plasmonic materials[13,24] to include the nonlinear optical absorption case. Specifically, we derive the third-order nonlinear susceptibilities and calculate the nonlocal nonlinear optical absorption coefficient $\beta$ as a function of the film thickness and angle of incidence of the incoming light beam, to show full consistency of our generalized theory predictions with unique peculiarities of nonlinear optical absorption enhancement observed experimentally for ultrathin TD films.

## Results

**Physical and optical properties of TiN films**

For this study, TiN thin films with varying thicknesses and Al-alloyed TiN films with different Al compositions were epitaxially grown on *c*-plane sapphire substrates by PA-MBE[49-52]. The thickness of TiN films ranges from 2 nm to 50 nm (nominal values of thickness are adopted), while that of Al-alloyed $Ti_{1-x}Al_xN$ films increases with Al composition, from ~60 nm at $x = 0.2$ to ~160 nm at $x = 0.6$. Figure 1a shows top-view scanning electron micrographs (SEM) of 4 nm and 50 nm thick TiN films, revealing smooth surfaces and high surface coverage even for the 4 nm epilayer, along with a densely packed crystalline structure for the 50 nm film. The 4 nm thick TiN film exhibits a low root-mean-square (RMS) surface roughness of ~1.5 Å as shown in the inset of Fig.1a, measured by atomic force microscopy (AFM) under a scan area of 2×2 $\mu m^{51,52}$. These results confirm that the impact of surface roughness on the ENZ effect is minimal in our MBE-grown ultrathin TiN films[53]. The structural properties of samples are investigated using X-ray diffraction (XRD) measurement. Figure 1b shows the XRD curves of the 4 nm and 50 nm thick films along with that of $Ti_{0.8}Al_{0.2}N$. The peaks at 36.7° and 41.7° correspond to TiN (111) and *c*-plane sapphire (0006), respectively. The XRD peak-width of the 50 nm thick TiN film is ~0.10°, indicating excellent crystalline quality. The broadened peak observed for the 4 nm thick TiN film is attributed to the reduced crystalline size associated with decreasing thickness. The satellite interference fringes (Pendellösung fringes) in the XRD curves are used to accurately estimate the film thickness through their fringe periods. In $Ti_{0.8}Al_{0.2}N$, the replacement of Ti atoms with smaller Al atoms leads to a slight shift of the TiN diffraction peak, and the magnitude of this shift increases with higher Al incorporation[51].

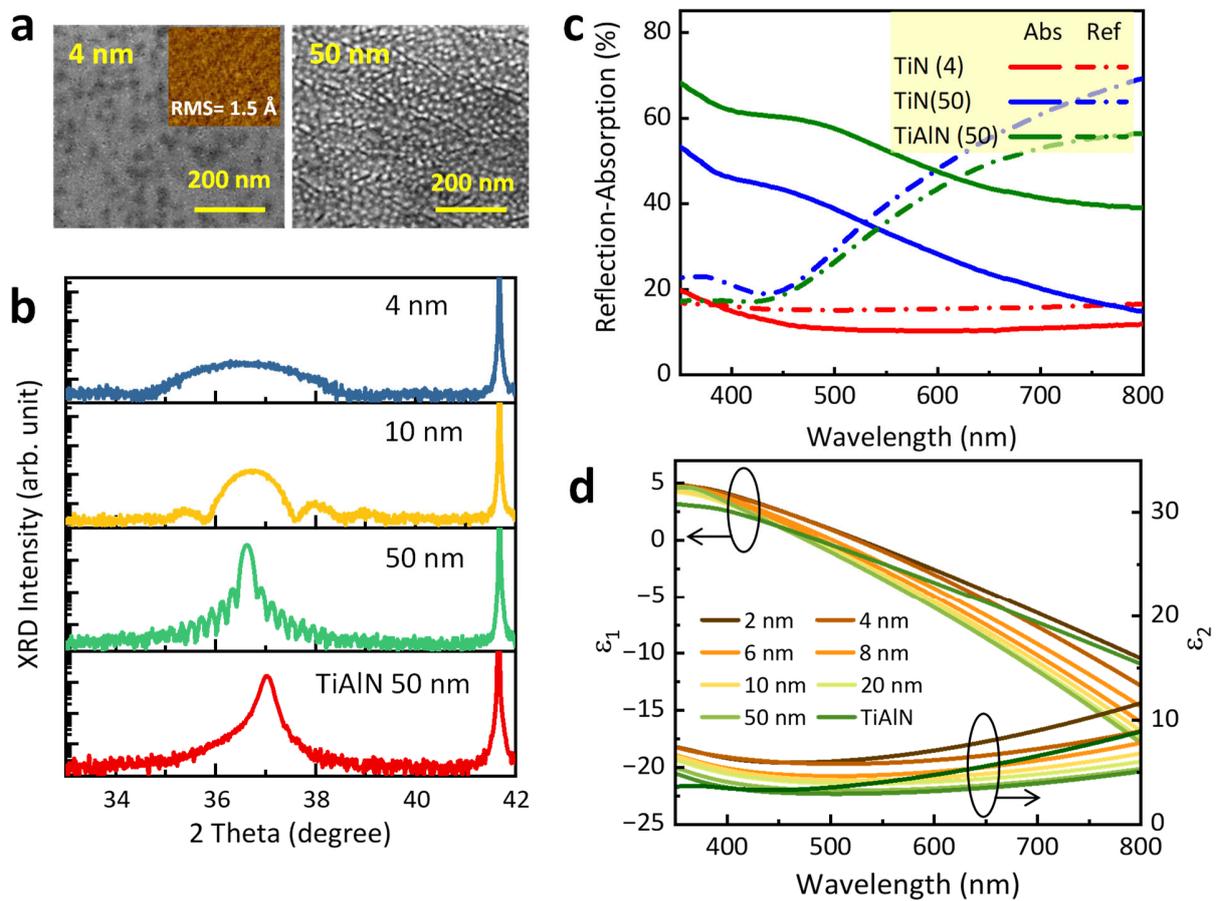

**Figure 1.** Crystalline and optical properties of TiN epitaxial films on *c*-plane sapphire. (a) The SEM images of the 4 nm and 50 nm thick TiN films. Inset shows the AFM image of the 4 nm thick TiN with RMS roughness of 1.5 Å. (b) XRD curves (X-ray wavelength: 1.54 Å) of TiN (4 nm and 50 nm) and $Ti_{0.8}Al_{0.2}N$ films. The peaks at 36.7° and 41.7° correspond to TiN (111) and *c*-plane

sapphire (0006). (c) Comparison of the reflection and absorption spectra for the 4 nm and 50 nm thick TiN films and 50 nm thick $Ti_{0.8}Al_{0.2}N$ film. (d) Real and imaginary parts ($\varepsilon_1$ and $\varepsilon_2$) of the dielectric functions of TiN and $Ti_{0.8}Al_{0.2}N$ films in the visible spectral range.

Figure 1c shows the optical absorption (solid curves) and reflection (dashed curves) spectra of 4 nm and 50 nm thick TiN films together with those of 50 nm thick $Ti_{0.8}Al_{0.2}N$ film, all in the visible range. While the thick TiN and $Ti_{0.8}Al_{0.2}N$ films exhibit a high reflection onset above ~450 nm in wavelength, corresponding to the interband transition, the 4 nm thick TiN film appears nearly transparent[50]. Notably, despite the small amount of Al composition, the $Ti_{0.8}Al_{0.2}N$ film exhibits degraded crystallinity[51] and markedly higher optical absorption compared with the 50 nm thick TiN film, resulting in a substantial increase in optical loss.

Figure 1d shows the dispersion of the real and imaginary parts of the linear dielectric response functions ($\varepsilon = \varepsilon_1 + i\varepsilon_2$) of the TiN films of varied thicknesses along with those of the 50 nm thick $Ti_{0.8}Al_{0.2}N$ film, measured by spectroscopic ellipsometry. In the metallic regime, all $\varepsilon_1$ functions exhibit negative values at longer wavelengths, then cross zero and become positive at wavelengths corresponding to the ENZ wavelength of each film. For the 50 nm thick TiN film, the ENZ wavelength is located at around 465 nm, and it continuously redshifts towards longer wavelengths (smaller frequencies) as the films get thinner[50]. Note that in our previous work, the screened plasma frequency ($\omega_p^{3D}$) of bulk (~100 nm thick) TiN fabricated by MBE method was measured to be 2.8 eV (=442 nm wavelength)[49], which is significantly greater than the values reported for thin and ultrathin TiN films by other experimental studies[26,54,55], thus indicating that the film thickness is an important parameter to control the plasmonic properties of the film[13,24]. Figure 1d also exhibits that $\varepsilon_2$ for the 50 nm thick TiN film is small and nearly nondispersive across the visible range. On the contrary, both thinner and Al-doped TiN films show higher $\varepsilon_2$ values than their thicker counterpart (50 nm thick TiN), indicating that thickness reduction (or alloy-induced compositional disorder[51]) unavoidably leads to an overall optical loss increase. More details of the optical properties of ultrathin TiN films and Al-doped TiN films, including their thickness- and composition-dependent behavior, can be found in our previous reports[50-52].

**Z-scan measurement for nonlinear optical properties**

The nonlinear optical properties of the ultrathin TiN films with various thicknesses were investigated using the open-aperture (OA) form of the Z-scan technique[56,57]. During these measurements, a typical laser intensity of ~5–40 GW/cm$^2$ at the focus was applied to samples at various incident angles. The scanning range of samples along the direction of light propagation (z-direction) is defined to be $z/z_0$, where $z_0 (= \pi w_0^2/\lambda)$ is the Rayleigh length and $w_0$ is the waist radius of Gaussian beam[57]. The closed-aperture (CA) Z-scan measurement for determining the nonlinear refraction could not be performed because of poor signal-to-noise ratio. Reported nonlinear optical parameters of metallic materials often exhibit significant variation due to differences in excitation wavelength, pulse duration, and light intensity (see **Table S1**). In particular, long-pulse excitation can cause heat accumulation in metals, complicating accurate evaluation of their intrinsic nonlinear responses using the Z-scan technique. To ensure that thermal effects do not influence our measurements, we first examine the ultrafast transient behavior of the TiN films. Figure 2a presents the transient reflectivity responses of the 6 nm thick and 40 nm thick TiN films along with that of gold, all measured at 500 nm. Additional information on the ultrafast carrier dynamics of TiN is available in our previous work[51]. Our TiN films exhibit significantly stronger and faster electron-phonon coupling compared to gold. Following the sub-picosecond electron-phonon energy transfer, phonon relaxation in our TiN occurs within ~1 ns, whereas it persists much longer in Au. Importantly, this relaxation time is far shorter than the 1 kHz repetition rate of our femtosecond laser system, preventing heat accumulation between consecutive pulses. Furthermore, **Figure S1** shows that the lattice temperatures of both ultrathin (6 nm) and thick (40 nm) TiN films, calculated using a two-temperature model (TTM), remain below 400 K under our excitation conditions. The electron-phonon coupling time $t_{ep}$ ($\approx$ 93 fs) used in the TTM calculations is extracted from exponential decay fits to the transient reflectivity ΔR/R of the TiN films in Fig. 2a. These results confirm that the Z-scan measurements on refractory TiN layers are free from thermal artifacts[58]. Additional details of the TTM calculations are provided in **Supplementary Information S1**.

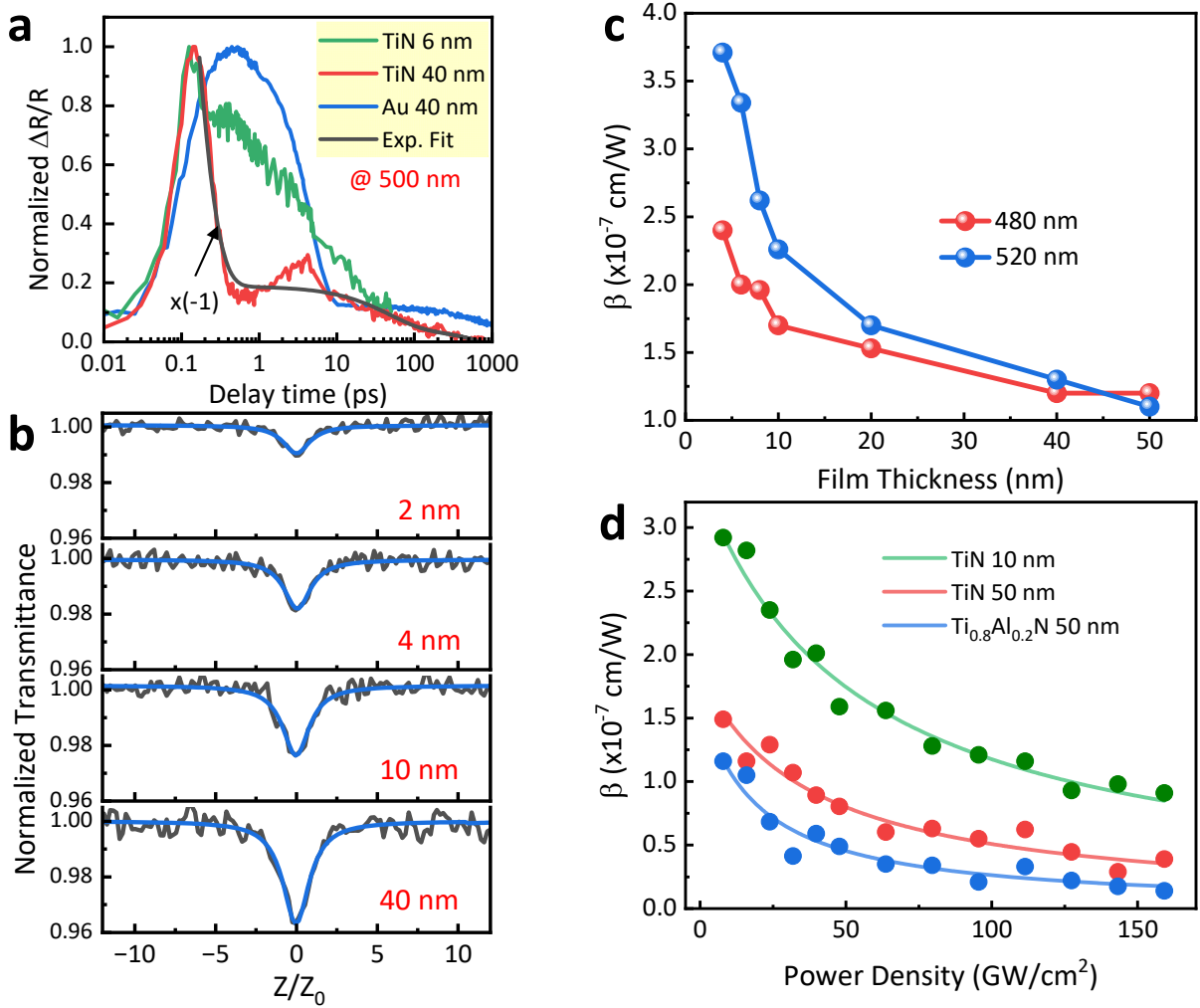

**Figure 2.** Transient reflectivity and Z-scan measurements. (a) Temporal evolution of normalized ΔR/R of 6 nm thick and 40 nm thick TiN films at 500 nm in a long delay time scale. (b) OA Z-scan traces of the TiN films of varied thicknesses. Blue solid lines show the best-fit curves with fitting parameters adjusted for each film individually. (c) The thickness dependence of the nonlinear absorption coefficient $\beta$ measured at wavelengths of 480 and 520 nm. (d) The light intensity dependence of $\beta$ for TiN (10 and 50 nm thick) and $Ti_{0.8}Al_{0.2}N$ (50 nm thick). Solid curves depict the saturation behavior of the $\beta$ coefficient $[= \beta_0/(1 + I/I_s)]$.

Figure 2b shows the Z-scan signals for TiN epilayers with thicknesses ranging from 2 nm to 40 nm, measured at a wavelength of 480 nm, close to the ENZ wavelength of the thick TiN film. For all samples, the transmitted Z-scan signal decreases as the sample approaches the focal plane, characteristic of a reverse saturable absorption (RSA) response. This behavior is attributed to various nonlinear absorption processes such as free-carrier absorption, two-photon absorption, and excited-state absorption. Similar RSA behavior has also been reported for gold and silver near their plasmon resonance frequencies[58,59]. Meanwhile, the previously reported saturable absorption (SA) hump due to ground state plasmon bleaching, mainly associated with defects[60], was not observed in our MBE-grown TiN epilayers. No obvious Z-scan signal was observed from the bare sapphire substrate.

We apply the conventional Z-scan data analysis method to obtain the nonlinear optical parameters of materials[57,61,62]. Upon intense laser excitation, the transmitted light intensity ($I$) through the material can be described by

$$\frac{dI}{dz} = -[\alpha_0 + \beta(I)I]I, \quad \beta(I) = \frac{\beta_0}{1 + I/I_s} \quad (1)$$

Here, $\alpha_0$ is the linear absorption coefficient and $\beta(I)$ is the intensity-dependent nonlinear absorption coefficient, where $\beta_0$ represents the low-intensity nonlinear absorption and $I_S$ denotes the saturation intensity. Combining with the Gaussian intensity profile of the laser beam, the solid curves in Fig. 2b are obtained by the least-square fit to the normalized OA Z-scan data using the relation[63]

$$T_{\text{OA}}(z) = 1 - \frac{q_0}{2\sqrt{2}\left(1 + z^2/z_0^2\right)} \qquad (2)$$

with the propagating distance z and $q_0 = \beta I_0 L_{\text{eff}}$, where $L_{\text{eff}}$ is the effective size of the sample in the direction of propagation calculated as $L_{\text{eff}} = \left(1 - e^{-\alpha_0 d}\right)/\alpha_0$ and $d$ is the thickness of the sample. The linear absorption coefficient $\alpha_0$ varies with film thickness and its dependence on thickness is shown in **Figure S2**.

Figure 2c presents the thickness dependence of the $\beta$ retrieved from the analysis of OA Z-scan data measured at wavelengths of 480 nm and 520 nm, close to the ENZ wavelengths of the thick and ultrathin TiN films, respectively. For thick TiN films, the average $\beta$ value is ~$(1.3\pm0.5)\times10^{-7}$ cm/W, nearly an order of magnitude larger than the RSA-induced $\beta$ reported for TiN nanoparticles at wavelength of 600 nm[44]. The nonlinear absorption coefficient increases sharply as the film thickness decreases, and the responses of films in the TD thickness range measured at the 520 nm wavelength are notably stronger than those measured at 480 nm, since this wavelength is closer to the ENZ wavelength of these ultrathin films. For instance, the $\beta$ of the 4 nm thick film measured at 520 nm is roughly three times greater than that of the 50 nm film. This behavior is comparable to the 4- to 5-fold enhancement reported for gold films of varied thickness, indicating a similar trend of nonlinear amplification in the TD thickness regime[58].

Figure 2d further illustrates the relationship between nonlinear optical absorption and excitation laser intensity for the 10 nm and 50 nm TiN films as well as the 50 nm $Ti_{0.8}Al_{0.2}N$ film, all measured at a fixed wavelength of 480 nm. The optical nonlinearity of the 50 nm TiN film is significantly less than that of the 10 nm TiN film, but still greater than that of the $Ti_{0.8}Al_{0.2}N$ film of the same thickness. As the excitation intensity increases, the positive $β$ coefficient (attributed to RSA) gradually decreases and eventually reaches saturation for all samples. The solid curves in Fig. 2d are obtained from Eq.(1) using $β_0$ values of 3.4, 1.9, and 1.7×10$^{-7}$ cm/W and saturation intensities $I_s$ of 53.3, 37.6, and 18.8 GW/cm$^2$ for the 10 nm TiN, 50 nm TiN, and 50 nm $Ti_{0.8}Al_{0.2}N$ films, respectively. No laser-induced surface damage was observed even for the 10 nm TiN film under this high optical illumination, in sharp contrast to ultrathin sputtered gold films[58], with a very low radiative damage threshold of only 0.2–0.4 GW/cm$^2$. Achieving strong Kerr-like optical nonlinearity requires both a high damage threshold and a pronounced saturation effect[18]. Smaller $β$ and $I_s$ of the $Ti_{0.8}Al_{0.2}N$ film as compared to those of TiN films can be attributed to the optical absorption variation due to Al doping. Our results, combined with the refractory nature of TiN material, stress its advantages as an ideal TD plasmonic material for high-intensity nonlinear optical applications.

**Nonlinear optical properties: Incident angle dependence**

The enhancement of optical nonlinearity in ENZ materials is known to exhibit characteristic dependences on the polarization and incident angle of the excitation beam[20]. Figures 3a and 3b show the Z-scan signals of the 10 nm thick TiN film measured with the p- and s-polarized 480 nm excitation beams, respectively, for angles of incidence ($θ$) varying from 0° to 70° (with a schematic shown in the top inset). The p-polarized response varies widely with $θ$, whereas the s-polarized response is less sensitive to the angle of incidence. To further illustrate the enhancement effect for the $\beta$ coefficient, we measure the angle-dependent Z-scan responses of TiN films with varying thicknesses. Figure 3c shows the p-polarized $\beta(\theta)$ coefficients for the 4 nm, 6 nm, 8 nm, and 50 nm TiN films obtained from the Z-scan signals at the wavelength of 480 nm. For the 50 nm film, the $\beta$ coefficient slightly increases with $\theta$ and then gradually decreases after around 10°. In contrast, for the ultrathin films, the $\beta$ coefficient increases substantially with the angle of incidence and reaches a maximum value at $\theta_{max}$, which shifts toward larger angles as the film thickness decreases. For the 4 nm thick film at $\theta_{max} \sim 45°$, in particular, the $\beta$ coefficient is about three times greater than its value at 0°, and nearly six times greater than $\beta(45°)$ for the 50 nm thick film. This enhancement observed in ultrathin TiN films is modest compared to that of TCOs, however, the characteristic polarization and angle-of-incidence dependences are consistent with the excitation of ENZ plasma modes predicted previously for ultrathin metals[12,13,21,64].

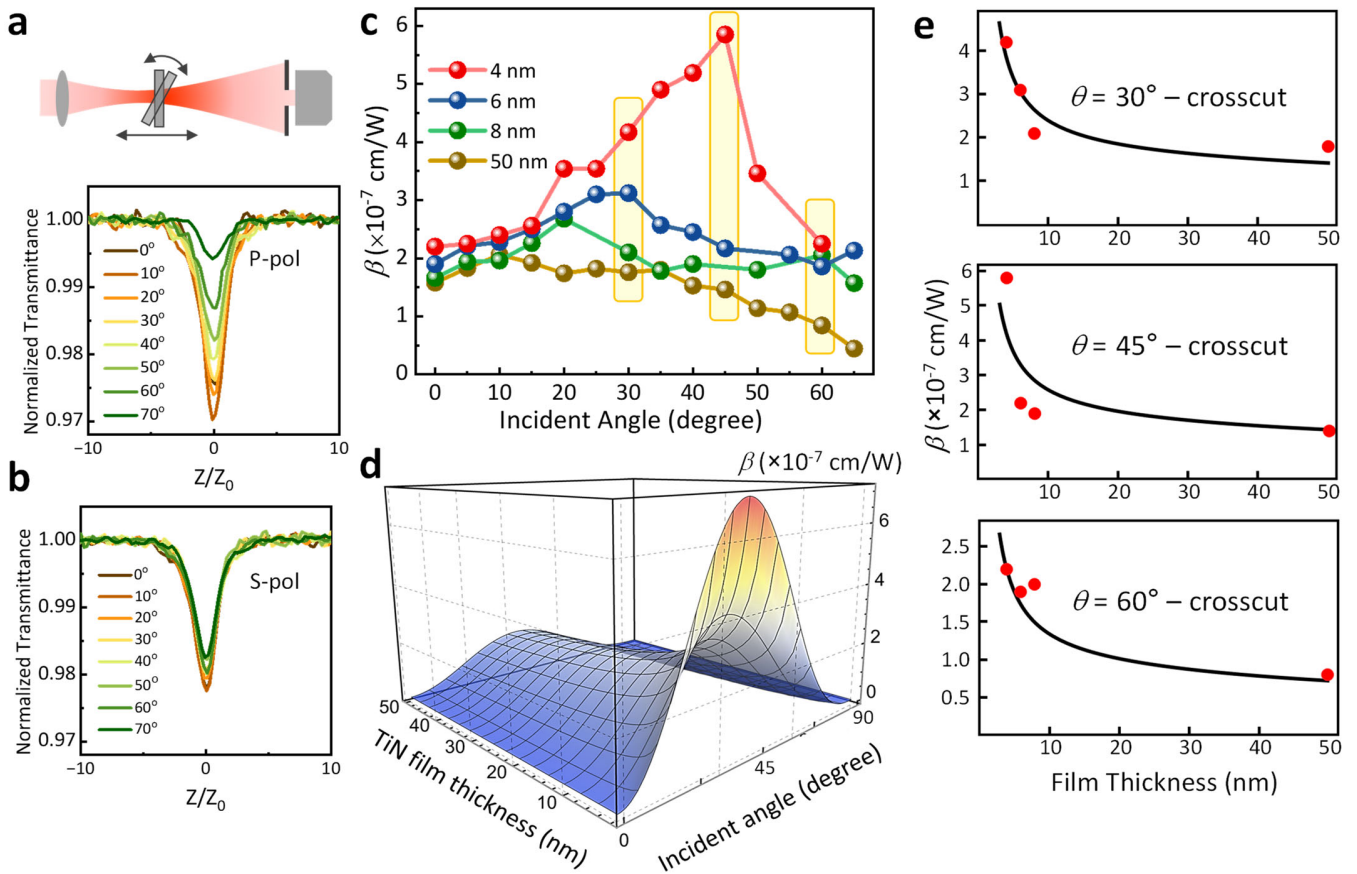

**Figure 3.** (a,b) $\theta$-dependent OA Z-scan traces of the 10 nm thick TiN film measured with p- and s-polarized excitation beams, respectively. The top inset in (a) shows the schematic of the $\theta$-dependent measurement setup. (c) Summary of the $\beta(\theta)$ dependences measured at 480 nm for TiN films with different thicknesses $d$. Ultrathin epilayers ($d$ <10 nm) have the maximum value of $\beta$ at oblique incident angles. Vertical rectangles highlight the data points presented in (e) to compare with theory. (d) Theoretical $\beta$ nonlinear optical absorption coefficient as a function of $d$ and $\theta$ as given by Eqs.(3)-(6) for the TiN films deposited on the sapphire substrate in air, with $\omega=\omega_p^{3D}$ (corresponding to the 480 nm wavelength), $\Gamma_{bulk}=20$ meV, and $C=4.4\times10^5$ (both chosen by adjustment to match the measurements). (e) Crosscuts of the $\beta(d,\theta)$ surface in (d) by the vertical planes of constant $\theta$ (=30°, 45°, and 60°) along with the experimental data points (red dots) for films of $d=4, 6, 8$, and 50 nm.

This behavior can be understood by considering that films thinner than 10 nm can no longer be described by the local in-plane EM response. Instead, due to the strong vertical electron confinement, they exhibit confinement-induced (and so thickness-dependent) nonlocal in-plane EM properties typical for TD materials[24,30,31]. Therefore, to account for the observed angular and thickness dependences of the $\beta$ nonlinear absorption coefficient, we use the concept of confinement-induced nonlocality occurring in TD plasmonic materials[13,24,34]. The linear nonlocal EM response theory of the TD plasmonic films is built on the Keldysh-Rytova (KR) pairwise electron interaction potential[65,66]. The KR potential accounts for the thickness-dependent vertical electron confinement in a thin plasmonic film sandwiched between dielectrics of permittivities $\varepsilon_{ss}$ and $\varepsilon_{sb}$, (superstrate and substrate, respectively) that are less than the in-plane background permittivity $\varepsilon_b$ of the film itself. As the film thickness decreases and becomes less than the typical in-plane distance between a pair of electrons confined inside, their repulsive KR interaction potential is mostly contributed by the region outside of the film due to the strong vertical confinement. Such dielectric confinement enhances the KR potential relative to that in a homogeneous medium of the same $\varepsilon_b$. This brings the plasmonic film into the TD regime, where the dimensionality of the system is reduced from 3D to 2D effectively, but there is still one and only remnant of the z-direction—the film thickness that comes as a product of $d$ and in-plane electron momentum $k$, thus making the EM response of the TD film nonlocal. More precisely[24,26], the KR electron-electron interaction potential leads to the nonlocal plasma frequency $\omega_p(k)$ that generalizes the local bulk plasma frequency $\omega_p^{3D}$ of the standard Drude model of normal metals. For thicker films, the KR potential turns into standard Coulomb potential, whereby the film thickness $d$ becomes an effective parameter to control the degree of nonlocality of the linear EM response and related EM properties of the TD plasmonic films[26-38].

For the ultrathin finite-thickness plasmonic TD film, the $\beta$ nonlinear optical absorption coefficient is not only a function of $\omega$ but also of $d$ and $\theta$. This function can be written in terms of the third-order nonlinear optical susceptibility $\chi^{(3)}(d,\theta,\omega)$ as follows[4,67]

$$\beta(\omega) = \frac{48\pi\omega}{10^{-7}c^2 n_0^2} \operatorname{Im} \chi^{(3)}(\omega) \qquad (3)$$

(Gaussian units), where $c$ is the velocity of light and $n_0$ is the film's refractive index determined by the in-plane nonlocal linear EM response function $\varepsilon(k,\omega)$ as follows[24,42]

$$n_0 = \sqrt{|\varepsilon(k,\omega)|}, \ \varepsilon(k,\omega) = \varepsilon_b \left\{ 1 - \frac{\omega_p^2(k)}{\omega(\omega + i\Gamma_D)} \right\}, \ \omega_p(k) = \frac{\omega_p^{3D}}{\sqrt{1 + \frac{\varepsilon_{ss} + \varepsilon_{sb}}{\varepsilon_b kd}}}, \ k = \sqrt{\varepsilon_{ss}} \frac{\omega}{c} \sin\theta. \quad (4)$$

Here, $\omega_p^{3D} = \sqrt{4\pi e^2 N_{3D}/(\varepsilon_b m)}$ is the standard expression for the screened plasma frequency of bulk TiN and $\Gamma_D = \Gamma_{bulk} + \Gamma_{surf}$ is the electron damping rate contributed by both bulk and surface (*d*-dependent) scatterings in the same way as it was earlier reported for similar transdimensional TiN films[26]. The single-frequency third-order nonlinear susceptibility is given by (see **Theoretical Methods**)

$$\chi^{(3)}(d,\theta,\omega) = -\frac{Ce^2}{m^2 d^2 (\omega_p^{3D})^4}\left\{1 + \frac{(\varepsilon_1+\varepsilon_2)\omega_p^{3D}}{\omega\sqrt{\varepsilon_{ss}\varepsilon_b}\,k_p^{3D} d\sin\theta}\right\}^2 \qquad (5)$$
$$\times \chi^{(1)}(d,\theta,\omega+i3\Gamma_D)\chi^{(1)}(d,\theta,\omega+i\Gamma_D)\left|\chi^{(1)}(d,\theta,\omega+i\Gamma_D)\right|^2$$

Here, $C$ is the dimensionless third-order nonlinearity strength constant and $\chi^{(1)}$ is the linear plasmonic susceptibility defined for incident *p*-polarized laser beam as $\chi^{(1)}(\omega) = \{\varepsilon(\omega) - \varepsilon_b\}\cos\theta$ per the standard rule[4], which in view of Eq.(4) leads explicitly to

$$\chi^{(1)}(d,\theta,\omega+i\Gamma_D) = -\frac{\varepsilon_b(\omega_p^{3D})^2\cos\theta}{\left\{\omega + \frac{(\varepsilon_{ss}+\varepsilon_{sb})\omega_p^{3D}}{\sqrt{\varepsilon_{ss}\varepsilon_b}\,k_p^{3D} d\sin\theta}\right\}(\omega+i\Gamma_D)} \qquad (6)$$

with $k_p^{3D} = \sqrt{\varepsilon_b}\,\omega_p^{3D}/c$ representing the bulk plasmon wave vector. This can be seen to acquire a non-trivial dependence on $d$ and $\theta$ for ultrathin plasmonic films. More specifically, it goes as $\chi^{(1)} \sim d\sin(2\theta)$ and traverses the maximum at $\theta_0 = \pi/4$ whenever the inequality

$$d\sin\theta \ll \frac{(\varepsilon_{ss}+\varepsilon_{sb})\omega_p^{3D}}{\sqrt{\varepsilon_{ss}\varepsilon_b}\,k_p^{3D}\omega} \qquad (7)$$

is fulfilled, which for non-zero $\theta$ can be seen being the case for small enough $d$ at all fixed frequencies. Moreover, one can see that Eq.(7) becomes even stronger for decreasing $\omega \leq \omega_p^{3D}$, where the equality corresponds to the ENZ regime of bulk plasmonic materials (or relatively thick plasmonic films).

Figure 3d shows the calculated nonlinear absorption coefficient $\beta$ as a function of $d$ and $\theta$ with $\omega$ fixed at $\omega_p^{3D}$, corresponding to the 480 nm laser wavelength used in our experiments. It can be seen from Eq.(4) that this is the onset of the ENZ regime (pertinent to thick conventional films) as $\omega_p$ can be seen to redshift (decrease) with $d$ reduction, in general—the tendency earlier confirmed experimentally[26,27]. We used Eqs.(3)-(6) in our calculations for TiN films on sapphire ($\varepsilon_{sb}$ = 9.2) in air ($\varepsilon_{ss}$ = 1), with $\Gamma_{bulk}$(TiN) = 20 meV and $C$ = 4.4×10$^5$ (both constants were adjusted manually to match our measured data points). The surface contribution $\Gamma_{surf}$ to the total damping rate $\Gamma_D$ in Eqs.(4) and (6) was simulated in the same way as it was previously reported for similar thickness TiN films[26]. The surface $\beta(d,\theta)$ in Fig. 3d shows the key behavior that supports our experimental observations. Specifically, the pronounced increase the $\beta$ coefficient can be seen with decreasing film thickness and its noticeable enhancement at larger oblique incident angles. Figure 3e shows the crosscuts of (d) by the vertical planes of constant $\theta$ (= 30°, 45°, and 60°), together with our data points measured (red dots) for films of $d$ = 4, 6, 8, and 50 nm (highlighted with vertical rectangles in Fig. 3c). The general agreement with the calculated values (solid curves) confirms the six-fold enhancement of nonlinear absorption at $\theta$ = 45° for the 4 nm film relative to the 50 nm film, as well as the observed higher absorption at $\theta$ = 30° as compared to $\theta$ = 60° for all film thicknesses.

**Nonlinear optical properties: Wavelength dependence**

Finally, based on the thickness-dependent $\theta_{max}$ values summarized in Fig. 3c, we examine the wavelength dependence of $\beta(\theta_{max})$ for each film. Figures 4a and 4b show the Z-scan responses of the 6 nm and 50 nm TiN films at different wavelengths within their respective ENZ spectral windows. Both films exhibit persistent RSA behavior across the entire wavelength range, a trend consistently observed for all investigated thicknesses. In contrast, the Z-scan signals of the 50 nm Ti$_{0.8}$Al$_{0.2}$N and 90 nm Ti$_{0.6}$Al$_{0.4}$N films (Figs. 4c and 4d) show a clear transition from RSA to SA around 490-510 nm.

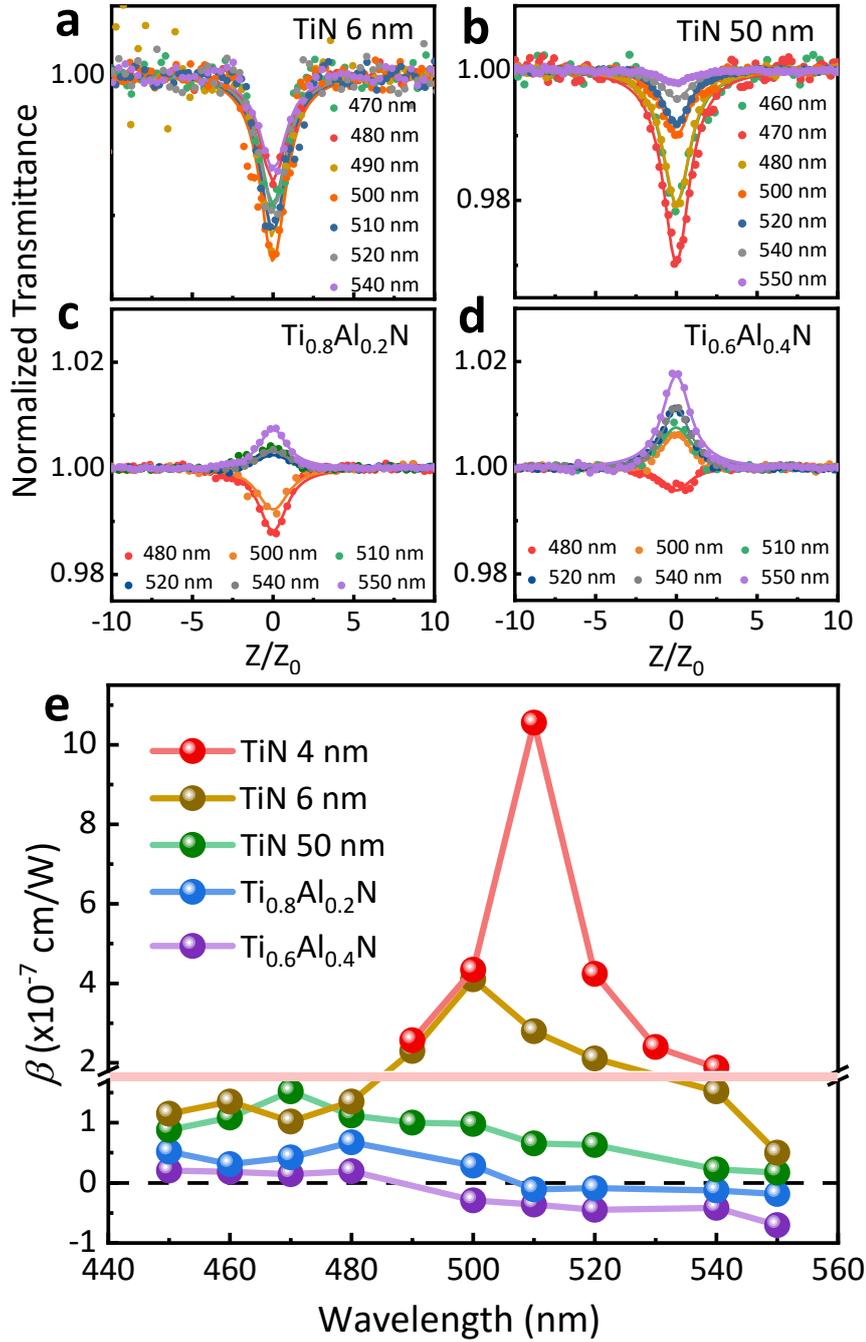

**Figure 4.** (a,b) OA Z-scan traces of the 6 nm and 50 nm TiN films measured in the ENZ spectral range. (c,d) OA Z-scan traces of $Ti_{1-x}Al_xN$. Transition from RSA to SA feature can be observed only in $Ti_{1-x}Al_xN$. (e) Wavelength dependence of the $\beta(\theta_{max})$ coefficient for the 4 nm, 6 nm, and 50 nm TiN films, as well as the 50 nm $Ti_{0.8}Al_{0.2}N$ and 90 nm $Ti_{0.6}Al_{0.4}N$ films (see Fig. 3c for $\theta_{max}$).

Figure 4e shows the wavelength dependences of $\beta(\theta_{max})$ for all film thicknesses, where positive and negative $\beta$ values correspond to RSA and SA behaviors, respectively. For the 50 nm thick TiN film, the $\beta$ coefficient exhibits a modest peak near 470 nm, close to its ENZ wavelength. However, as the films become thinner, the peak of the $\beta$ coefficient shifts to longer wavelengths, appearing at the 500 nm and 510 nm wavelengths for the 6 nm and 4 nm thick TiN films, respectively, and increases dramatically in magnitude. Notably, the maximum $\beta$ value of the 4 nm thick TiN film reaches ~$1.05\times10^{-6}$ cm/W, approximately seven times larger than that of the 50 nm thick film at 470 nm and over sixty times greater than the 50 nm film at 550 nm. This remarkable result is consistent with the expected ENZ-driven enhancement of optical nonlinearity in TD plasmonic films. Its origin can be understood from the inequality in Eq.(7), which, as discussed above, becomes more pronounced with decreasing $\omega$ (or with increasing wavelength). In this case the peak of the $\beta(\theta_{max})$ should necessarily shift towards longer wavelengths for thinner films.

For the Al-alloyed films, the wavelength dependence of $\beta(\theta_{max})$ differs significantly from that of ultrathin TiN films. As shown in Fig. 4e, whereas TiN exhibits a sharp $\beta$ enhancement near its ENZ wavelength, the positive $\beta$ values of $Ti_{1-x}Al_xN$ instead cross over into the negative region around 500 nm wavelength. Typically, the RSA behavior of Z-scan signals arises from dominant excited-state absorption over ground-state bleaching. In our previous work[51], we found that, in addition to degraded crystalline quality, the number of available excited states in $Ti_{1-x}Al_xN$ alloy decreases with increasing Al concentration, which could weaken the RSA response and change the underlying nonlinear absorption mechanism[68,69]. Furthermore, as shown in Fig. 1c, $Ti_{1-x}Al_xN$ exhibits stronger absorption than TiN, which further increases with higher Al content[51]. Such an additional loss is expected to suppress the excitation of ENZ modes in $Ti_{1-x}Al_xN$, emphasizing the critical role of high crystallinity and low-loss metallic quality in achieving pronounced ENZ effects.

Combined with the angle-dependent enhancement of the *β* coefficient discussed above, the distinct wavelength dependence near the ENZ region provides compelling experimental evidence for ENZ-induced nonlinear amplification in refractory ultrathin TiN films, despite longstanding challenges associated with optical losses and extreme thickness constraints.

**Discussion**

In this work, we experimentally demonstrate the ENZ-induced optical nonlinearity enhancement in ultrathin TiN films operating in the TD thickness regime. The characteristic signatures of the ENZ effect are systematically studied by measuring the thickness-, wavelength-, and incident angle-dependent nonlinear absorption coefficient *β* using the Z-scan technique. The stringent requirement for nanoscale metallic films to exhibit ENZ responses was met by fabricating high-quality ultrathin TiN films by means of molecular beam epitaxy, which ensures both excellent crystalline quality and precise thickness control. Although optical loss—an inherent limiting factor for ENZ excitation in metals—increases as the films become thinner, the confinement-induced nonlocality in TD films dominates, leading to a pronounced enhancement of optical nonlinearity with decreasing film thickness. This effect is particularly strong for ultrathin TiN films (4 nm thick in our study) illuminated by p-polarized radiation at incident angles close to 45° with wavelength close to the ENZ wavelength of the film. The excellent agreement between experimental results and the theoretical model developed for confinement-induced nonlocal nonlinear susceptibilities proves a behavior that is unique to TD plasmonic materials. Just as the theory predicts, wavelength-dependent measurements further reveal that the *β* peak redshifts as the films become thinner with ultrathin films exhibiting up to a sixty-fold enhancement compared to thicker films at non-ENZ wavelengths. The absence of such effects in $Ti_{1-x}Al_xN$ alloys underlines the importance of high crystallinity and low-loss plasmonic behavior for accessing ENZ-driven nonlinear amplification.

Our results establish TD films of TiN, a thermally robust, CMOS-compatible plasmonic material, as a compelling flexible TD material platform for realizing nanoscale nonlinear photonic devices enabled by ENZ plasmonics. The third-order optical nonlinearity enhancement demonstrated here opens new opportunities for all-optical, nanoscale-size nonlinear photonic devices operating in the visible optical range.

**Methods**

**Growth of ultrathin TiN and TiAlN epitaxial films.** The 2-inch wafer scale TiN films on double-side polished *c*-sapphire were grown by a nitrogen-plasma-assisted molecular-beam epitaxy system (DCA Instruments Oy, Turku, Finland). The thermal pre-treatment at 950 °C was applied for 2 hours for surface cleaning. The MBE base pressure was kept about $1 \times 10^{-10}$ Torr before growth. For the growth of Al-doped TiN, the Ti and Al fluxes were supplied and controlled by tuning the temperature of Knudsen cell, and the active nitrogen source was introduced via RF plasma gun (Veeco Instruments, U.S.A.). In-situ reflection high-energy electron diffraction pattern (RHEED) were observed during growth to monitor crystalline variation during TiN and $Ti_{1-x}Al_xN$ films growth.

**Optical Characterization System**

X-ray diffraction (XRD) *θ-2θ* scans (Bede D1 HR-XRD) of TiN films are performed using an XRD diffractometer with a monochromatic Cu K*α* radiation (λ = 1.5406 Å) to characterize the reflections along the surface normal and accurate determination of TiN film thicknesses. The elemental compositions of TiN and $Ti_{1-x}Al_xN$ were analyzed by X-ray photoelectron spectroscopy (XPS) spectra (PHI Quantera II, ULVAC-PHI Inc.) using Al K*α* X-ray (photon energy *hν* = 1486.6 eV) emission with a 100 μm diameter spot[50,51]. A low-energy (2 kV) Ar$^+$-ion was utilized to remove self-limited native oxide layers on the sample surface. The morphology of TiN films was studied by scanning electron microscopy (SEM, JSM-7000F) and AFM. The dielectric properties of TiN and TiAlN films were measured using a spectroscopic ellipsometer (m2000, J. A. Woollam Co.), which was supported by the National Nano Device Laboratories (NDL) in Taiwan. The nonlinear absorption coefficient $\beta$ was obtained via open-aperture (OA) Z-scan configurations. The thickness of our TiN films is much smaller than the Rayleigh length ($z_0$) of ~2 mm. An amplified femtosecond Ti:sapphire laser (Spitfire, Spectra Physics) providing <100 fs laser pulses at 1 kHz repetition rate was used as the excitation and probe light source for both Z-scan and ultrafast pump–probe measurements. All optical measurements were performed under ambient conditions with a humidity of 45 %.

**Theoretical Methods**

To understand and explain the measured angular dependences of the $\beta$ nonlinear absorption coefficient, we generalize the nonlocal linear EM response theory of TD plasmonic materials[13,24,34] to include the nonlocal nonlinear optical absorption case. We do this by using the straightforward connection between the linear and nonlinear EM response functions (susceptibilities) known as Miller's rule[4,70]. The rigorous mathematical relations between the linear and higher-order (2nd and 3rd) nonlinear optical susceptibilities (Miller's rule) were recently derived quite generally within the framework of the quantum anharmonic oscillator model[70]. There are only two intrinsic dimensional parameters in Eq.(4) representing the in-plane nonlocal linear EM response function of the TD plasmonic film system. They are thickness $d$ and plasma frequency $\omega_P(k)$, which when combined as $\omega_P(k)^2/d^{\,2}$ provide the proper dimension of cm$^{-2}$ s$^{-2}$ for the oscillator's 3rd-order anharmonicity parameter responsible for the 3rd-order nonlinear optical susceptibilities $\chi^{(3)}(\omega)$ and $\chi^{(3)}(3\omega)$ of nonlinear systems. With this in mind, we rewrite the relevant general single-frequency $\chi^{(3)}$-Miller's rule of Ref.70 expressing its dimensional pre-factor in terms of $\omega_p(k)$ and $d$, the key intrinsic characteristics (confinement-induced nonlocality and $d$-dependence) representative of the TD plasmonic film systems, which leads to Eq.(5).

## Acknowledgments

H.A. is supported by the National Science and Technology Council of Taiwan (NSTC 113-2112-M-A49-021-MY3). S.G. is supported by the National Science and Technology Council of Taiwan (NSTC 112-2112-M-007-049-MY3). I.V.B. acknowledges support from the US Army Research Office under award no. W911NF2310206. H.A. also acknowledges Dr. Yu-Jung Lu at the Research Center for Applied Sciences, Academia Sinica in Taiwan for providing access to the laser system.

## Author Contributions

H.A. planned the research, analyzed the experimental results, and wrote the manuscript. The laboratory of S.G. prepared TiN and TiAlN films and performed the physical property analysis. F.T. and T.K. performed the z-scan experiments. I.H. and Y.L. performed the optical measurement and carrier dynamics study. I.V.B. conducted the theoretical calculations and contributed to manuscript writing. All authors have given approval to the final version of the manuscript.

## Competing interests

The authors declare no competing interests.

## Additional information

**Supporting Information**. Two-temperature model (Figure S1) The linear absorption coefficient of TiN films obtained from the absorption spectra. (Figure S2) Comparison of the nonlinear optical absorption coefficient of various materials. (Table S1)

**Correspondence** and requests for materials should be addressed to Hyeyoung Ahn.

\* Tel) +886-3-5712121-56369, Fax) +886-3-5716631, E-mail) hyahn@nycu.edu.tw

**Data Availability**

The manuscript includes all data generated or analyzed during this study. The authors declare that all data supporting the findings of this study are available within the article and its Supplementary Information.

**Abbreviations**

Au, gold; TiN, titanium nitride; TiAlN, titanium aluminum nitride; TD, transdimensional; ENZ, epsilon-near zero; TCO, transparent conductive oxide; ITO, indium tin oxide; MBE, molecular-beam epitaxy; XRD, X-ray diffraction; SEM, scanning electron microscopy; SE, spectroscopic ellipsometry; OA, open aperture; RSA, reverse saturable absorption; SA, saturable absorption; e-ph, electron-phonon; KR, Keldysh-Rytova

Supporting Information for

# Confinement-Induced Nonlocality and Optical Nonlinearity of Transdimensional Titanium Nitride in the Epsilon-Near-Zero Region

*Fan-Ting Tseng,[1] I-Hung Ho,[1] Ting-Jui Kuo,[1] Shangjr Gwo,[2] Igor V. Bondarev[3], and Hyeyoung Ahn[1]\**

[1]Department of Photonics, College of Electrical and Computer Engineering, National Yang Ming Chiao Tung University, Hsinchu 30010, Taiwan.

[2]Department of Physics, National Tsing-Hua University, Hsinchu 30013, Taiwan.

[3]Department of Mathematics and Physics, North Carolina Central University, Durham, NC 27707, USA

\*Correspondence to: hyahn@nycu.edu.tw

Contents

Two-temperature model calculation. The linear absorption coefficient ($\alpha_0$) of TiN films obtained from the thickness dependent absorption spectra. Comparison of the nonlinear optical absorption coefficient of various materials.

## S1. Two-temperature model

The temporal dependence of the electron and lattice temperatures was calculated using the Two-Temperature Model. In TTM, the electron and lattice temperatures reach equilibrium according to the electron–phonon coupling coefficient ($G$) and the respective specific heat capacities of electrons ($C_e$) and the lattice ($C_l$)[1]. The TTM relates the temporal evolution of electron temperature ($T_e$) and lattice temperature ($T_l$) as

$$C_e(T_e)\frac{\partial T_e}{\partial t} = -G(T_e - T_l) + S(t) \qquad (1)$$

$$C_l(T_l)\frac{\partial T_l}{\partial t} = G(T_e - T_l) \qquad (2)$$

where $S(t)$ is the absorbed power from the optical femtosecond pulse. The electron heat capacity term adopts the Debye approximation that

$$C_e = \gamma T_e \qquad (3)$$

where $\gamma$ is the electron contribution to the heat capacity[2,3].

The source term in Eq. (1) can be described by following equation:

$$S(t) = \frac{\sqrt{\left(\frac{4\ln(2)}{\pi}\right)}(1 - R - T)F}{\delta t_p} \exp\left[-4\ln(2)\left(\frac{t - 2t_p}{t_p}\right)^2\right] \qquad (4)$$

where $R$ and $T$ are optical reflectivity and transmittance of the thin film, respectively, $\delta$ is the optical penetration depth, $t_p$ is laser pulse width, $F$ is laser pulse fluence.

Based on the electron-phonon coupling time ($t_{ep}$) from Fig. 2a, the value of the electron-phonon coupling coefficient, $G$ is calculated by using a simple relation

$$t_{ep}^{-1} \approx G\left(\frac{1}{C_e} + \frac{1}{C_l}\right) \qquad (4)$$

As a result, $G$ for TiN and is estimated to be $8.5\times10^{17}$ W/m$^3$ K with $\gamma=274$ J/m$^3$ K$^2$ and $C_l=3.13\times10^6$ J/m$^3$ K  Here, due to relatively small contribution of lattice, the electron heat capacity term dominantly contributes to calculation of $G$ in Eq. (4).

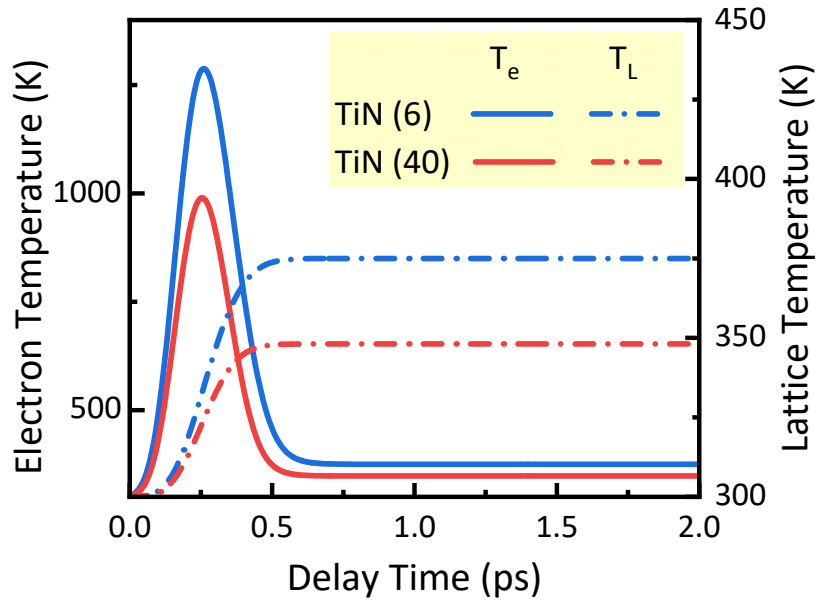

**Figure S1.** Temporal evaluation of electron and lattice temperatures calculated using the Two-Temperature Model for the 6 and 50 nm TiN films. In the calculation, the pump fluence of 50 fs laser pulses is assumed to be 10 GW/cm$^2$. The sub-picosecond electron-phonon coupling times are obtained from the transient reflectivity spectra shown in Fig. 2a.

## S2. Linear absorption coefficient of TiN films

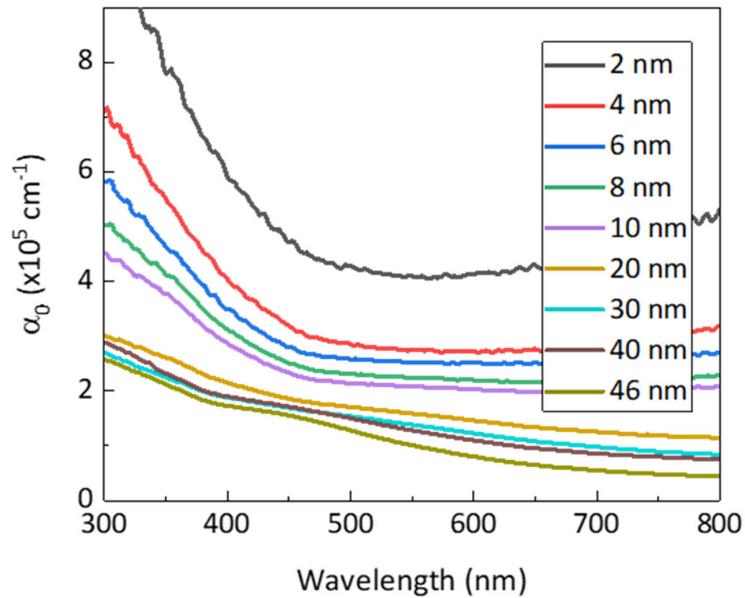

**Figure S2.** Linear absorption coefficient $\alpha_0$ calculated from the absorption spectra of TiN films with various thicknesses using the relation of $\alpha_0 = -\frac{\ln(1-A)}{d}$, where $d$ is the film thickness.

**Table S1. Comparison of the nonlinear optical absorption coefficient of various materials**

| Materials | Laser wavelength (nm) | Pulse width | $\beta$ (cm/W) | Sample preparing method | Ref. |
|---|---|---|---|---|---|
| ITO film | 1240 (ENZ) | 150 fs | $-7\times10^{-6}$ | commercial | 1 |
| AZO film | 1300 (ENZ) | 100 fs | $-2.5\times10^{-6}$ | PVD | 2 |
| Au (8.5 nm) | 532 | 35 ps | $-4.5\times10^{-3}$ | sputtering | 3 |
| Au (20 nm) | 630 | 6 ps | $6.7\times10^{-5}$ | evaporation | 4 |
| TiN film (52 nm) | 780 | 150 fs | $-6.8\times10^{-7}$ | PVD | 5 |
| TiN nanoparticles | 600 | 5 ns | $2.4\times10^{-8}$ | laser-ablation | 6 |
| TiN nanosheets | 530 (ENZ) | 35 fs | $-5.05\times10^{-6}$ | nitridation of $TiO_2$ | 7 |
| TiN nanosheets | 540 (ENZ) | 35 fs | $-3.78\times10^{-6}$ | nitridation of $TiO_2$ | 8 |
| TiN film (4 nm) | 510 (ENZ) | 50 fs | $1.7\times10^{-6}$ | MBE | This work |